\documentclass[conference]{IEEEtran}
\IEEEoverridecommandlockouts

\usepackage{cite}
\usepackage{amsmath,amssymb,amsfonts}
\usepackage{algorithmic}
\usepackage{graphicx}
\usepackage{textcomp}
\usepackage{tabularx}
\usepackage{caption} 
\usepackage{multirow}
\usepackage{multicol}
\usepackage{comment}
\usepackage[dvipsnames]{xcolor}
\usepackage{float}
\usepackage{tikz}
\usetikzlibrary{calc}
\usepackage{stackengine}
\tikzset
  {main node/.style=
    {rectangle,fill=white!5,draw,minimum size=0.5cm,inner sep=2pt}
  }

\def\BibTeX{{\rm B\kern-.05em{\sc i\kern-.025em b}\kern-.08em
    T\kern-.1667em\lower.7ex\hbox{E}\kern-.125emX}}
    
\usepackage{pgfplots}
\pgfplotsset{compat=1.3}
\pgfplotsset{
    discard if/.style 2 args={
        x filter/.code={
            \edef\tempa{\thisrow{#1}}
            \edef\tempb{#2}
            \ifx\tempa\tempb
                
            \fi
        }
    },
    discard if not/.style 2 args={
        x filter/.code={
            \edef\tempa{\thisrow{#1}}
            \edef\tempb{#2}
            \ifx\tempa\tempb
            \else
                
            \fi
        }
    }
}

\newcommand{\Field}[1]{\mathbb{#1}}

\newcommand{\Set}[1]{\mathcal{#1}}
\newcommand{\Operator}[1]{\mathit{#1}}

\newcommand{\Sseq}[2]{\{#1, \dots ,#2\}}

\newcommand{\Matrix}[1]{\mathbf{#1}}
\newcommand{\Vector}[1]{\pmb{#1}}
\newcommand{\norm}[1]{\lVert #1 \rVert}

\newcommand{\normOp}[1]{\norm{#1}_\mathrm{op}}

\newcommand{\Vect}{\text{Vec}}

\newcommand{\Real}{\Field{R}}

\newcommand{\sS}{\Set{S}}
\newcommand{\sX}{\Set{X}}

\newcommand{\sT}{\Real^{V \times M^2}}
\newcommand{\sQ}{\Real^{V}}

\newcommand{\sD}{\Set{D}}
\newcommand{\sW}{\Real^{M\times M}}

\newcommand{\sI}{\Real^{M\times M}}

\newcommand{\opf}{\Operator{f}}
\newcommand{\opfi}{\opf^{(i)}}
\newcommand{\opfh}{\hat{\opf}}

\newcommand{\opE}{\Operator{E}}
\newcommand{\opC}{\Operator{C}}

\newcommand{\opEX}{\opE(\sX, \cdot)}
\newcommand{\ope}{\Operator{e}}
\newcommand{\opH}{\Operator{H}}

\newcommand{\oph}{\Operator{h}}
\newcommand{\opv}{\Operator{v}}
\newcommand{\opD}{\Gamma}

\newcommand{\opL}{\Lambda}

\newcommand{\mA}{\Matrix{A}}
\newcommand{\mI}{\Matrix{I}}
\newcommand{\mE}{\Matrix{E}}
\newcommand{\mX}{\Matrix{X}}

\newcommand{\mU}{\Matrix{U}}
\newcommand{\mV}{\Matrix{V}}

\newcommand{\tX}{\tilde{\mX}}

\newcommand{\vy}{\Vector{y}}
\newcommand{\vf}{\Vector{f}}
\newcommand{\vz}{\Vector{\gamma}}

\newcommand{\hf}{\hat{\vf}}
\newcommand{\ty}{\tilde{\vy}}

\newcommand{\define}{\triangleq}

\newcommand{\psc}{AICC}

\newcommand\blankfootnote[1]{%

  \let\svthefootnote\thefootnote%
  \let\thefootnote\relax\footnotetext{#1}%
  \let\thefootnote\svthefootnote%

}

\begin{document}

\title{A Learning-Based Approach to Approximate Coded Computation}

\author{Navneet Agrawal$^{1*}$, Yuqin Qiu$^{2*}$, Matthias Frey$^{1}$, Igor Bjelakovic$^{3}$, Setareh Maghsudi$^{3,4}$, \\ Slawomir Stanczak$^{1,3}$, and Jingge Zhu$^{2}$ \\
        \small $^{1}$ Technische Universität Berlin, Germany \\
        \small $^{2}$ Department of Electrical and Electronic Engineering, University of Melbourne, Victoria, Australia \\
        \small $^{3}$ Fraunhofer Heinrich Hertz Institute, Berlin, Germany \\
        \small $^{4}$ Department of Computer Science, University of Tübingen, Germany
}

\maketitle

\blankfootnote{
$*$ These authors contributed equally to this work.\\
NA, MF, IB and SS acknowledge financial support by the Federal Ministry of Education and Research of Germany in the program of "Souverän. Digital. Vernetzt." Joint project 6G-RIC, project identification numbers: 16KISK020K and 16KISK030. Furthermore, they were supported by BMBF under grant 01DR21009 and by the German Research Foundation (DFG) within their priority program SPP 1914 ``Cyber-Physical Networking''. This work was also supported by the Berlin University Alliance within the excellence strategy of the German federal and state governments. JZ acknowledges the support by the Australian Research Council under Project DE210101497 and the UoM-BUA Partnership scheme.
}

\begin{abstract}
    Lagrange coded computation (LCC) is essential to solving problems about matrix polynomials in a coded distributed fashion; nevertheless, it can only solve the problems that are representable as matrix polynomials. In this paper, we propose AICC, an AI-aided learning approach that is inspired by LCC but also uses deep neural networks (DNNs). It is appropriate for coded computation of more general functions. 
    Numerical simulations demonstrate the suitability of the proposed approach for the coded computation of different matrix functions that are often utilized in digital signal processing.

\end{abstract}

\section{Introduction}

Coded Computation (CC) refers to a class of distributed computation schemes in which an encoder injects redundancy into the input data before it is distributed to the worker machines to perform the desired computation task. This allows reconstruction of the full computation result with a decoding step even if not all the workers have returned computation results. Therefore, CC schemes can reduce latency and increase reliability in scenarios in which the worker nodes are, e.g., subject to random failures or straggling~\cite{Li2016}. 

The CC schemes with the recovery of the exact computation results are, for instance, applicable to matrix-matrix multiplication~\cite{yu2017polynomial,dutta2019optimal} and the computation of matrix polynomials~\cite{yu2019lagrange,soleymani2021analog}. Some papers have also proposed CC schemes for these problems that can work with a significantly lower number of worker results in the decoding step. This comes at the cost of recovering the computation result only approximately with a certain prescribed accuracy~\cite{chang2019random,charalambides2021approximate,jahani2021codedsketch}. Another line of research considers schemes that have a soft recovery threshold in the sense that they do not need a minimum number of worker results to obtain an estimate. Instead, they obtain estimates with increasing accuracy as more and more of these results become available~\cite{gupta2018oversketch,kiani2022successive,jeong2021epsilon}. Approximate coded computation is relevant for many applications, including those pertaining to approximate machine learning (ML) tasks using deep neural network (DNN) models. 

Many existing papers propose schemes specifically for the training stage of DNNs that deal with erasures as well as with maliciously injected incorrect computation results (seminal works include~\cite{chen2018draco, amiri2019computation}).
References \cite{kosaian2018learning,kosaian2019parity,kosaian2020learning} apply CC to the inference stage of ML computation tasks, and follow the traditional CC paradigm that divides the scheme into the three stages of encoding, computing, and decoding. These papers propose methods for ``learning'' a suitable erasure code, wherein, one approach jointly trains an auto-encoder for encoding and decoding operations, whereas, in another, the computation corresponding to the encoded data is a specially trained neural network. In both approaches, not only the system is constrained to a single erasure, but the performance also degrades significantly with increasing number of inputs.

In this work, we aim to expand the class of functions compatible with the approximate CC schemes to include non-polynomial functions other than, but not excluding, the evaluations of the inference stage of neural networks. For numerical evaluations, we focus on the approximate CC computations of the following functions, which we believe to be of particular interest in the context of wireless communications:
computation of 
($P_1$) the eigenvalues,
($P_2$) the dominant eigenvector corresponding to the largest absolute eigenvalue,
($P_3$) the matrix exponential, and
($P_4$) matrix determinant
of a square matrix.

To the best of our knowledge, the CC schemes for these functions have not yet been proposed in the state-of-the-art research. Our approach is substantially inspired by the Lagrange Coded Computation (LCC)~\cite{yu2019lagrange}, but it extends the idea by using neural networks to achieve the following improvements:
\begin{itemize}
\setlength\itemsep{0.5em}
    \item The class of computable functions are no longer limited to the matrix polynomials. 
    \item The proposed scheme guarantees a recovery threshold, which is a design parameter.
        In particular, its choice does not depend on the number of computation inputs  directly.
    \item Approximate function computation allows for a judicious trade-off between the accuracy and the computation time.
\end{itemize}

\section{Notation and Preliminaries}  \label{sec:system}

We consider a system consisting of a \emph{master} and $N$ \emph{workers}. 
Given a function $\opf:\sD \to \sS$, where $\sD \subset \sI$ and $\sS$ is some finite dimensional Euclidean space,
the master is required to compute function values
at points gathered in the set $\sX \define \Sseq{\mX_1}{\mX_K}$, where $\mX_k \in \sD$ for all $k=1, \dots, K$.
For this purpose, the master can distribute the computations to the workers.

In coded computation, the master node takes the input data $\sX$ to generate $N$ encoded data $\tX_i, i=1,\ldots, N$. 
The worker $i$ performs computation on the encoded data $\tX_i$ and returns the result to the master.
The coded computation scheme allows the master to recover all desired function values $f(\mX_k)$, $k=1,\ldots, K$ 
upon receiving the computation results from any subset of $R \leq N$ workers. 
The smallest value of $R$ for which this is possible is called the \textit{recovery threshold} of the scheme. 
In this way, the CC schemes can mitigate the impact of straggling or failing worker.

The LCC~\cite{yu2019lagrange} scheme is a coded computation scheme for computing polynomial functions $\opf$, say, of degree $d$.
The functioning principle of the LCC can be summarized as follows (see~\cite{yu2019lagrange} for details):
The data is encoded to $N$ matries $\tX_i$, $i=1,\ldots, N$, 
each of which amounts to the evaluation of a Lagrange polynomial of degree $K-1$ at distinct scalars $\alpha_n$.
The worker $n\in\Sseq{1}{N}$ then computes the desired polynomial $\opf$ on the encoded data $\tX_n$.
The computation result $\opf(\tX_n)$ can be viewed as the evaluation of a composite polynomial of degree $(K-1)d$,
formed by the Lagrange polynomial and the polynomial function $\opf$ at a point $\alpha_n$.
Hence, to recover this polynomial, the master only requires $R = (K-1)d +1$ results from the workers.
The desired function values can then be obtained by evaluating the polynomial at suitable points.
Although LCC is optimal in terms of the recovery threshold, 
it suffers from two crucial limitations: 
(i) the function $\opf$ is restricted to be a polynomial of the input in LCC; and
(ii) the recovery threshold $R$ of LCC grows proportionally with the number of inputs $K$ and degree $d$ of the polynomial $\opf$.

The core motivation for this work is to explore the possibility of extending the idea of LCC to general functions, while ensuring a recovery threshold.
Roughly speaking, we replace both the encoding and the computation operations in LCC by deep neural networks.
Particularly, these operations are carefully designed such that the computation results from workers are evaluations of a certain polynomial (reminiscent of LCC or any polynomial-based CC scheme), hence only polynomial interpolation is needed for the recovery. In the sequel, we will discuss how our construction can overcome the limitations of LCC mentioned above to some degree. However, we point out that we only aim to recover the result \textit{approximately}, whereas LCC and many other coded computation schemes guarantees a perfect recovery (up to numerical stability issues in the recovery phase).

\section{AI-aided Coded Computation Scheme} \label{sec:AICC}

We propose an \emph{AI-aided coded computation} (\psc) scheme that utilizes a learning-based approach
for coded computation, applicable to a large class of functions including non-polynomials.
\psc~employs a system design that inherits the recovery-through-interpolation aspect of the LCC,
but ensures a fixed recovery threshold that does not directly depend on the number of inputs or polynomial degree of the function. Note that although \psc~can only provide an approximate solution in general, in Section \ref{sec:simulations}, we show that our design achieves a sufficient level of accuracy. Since an approximate solution is good enough in many practical applications, \psc~is relevant in a wide range of applications.
In the following, we first describe the system architecture and then the training procedure of the proposed scheme.

\subsection{System architecture and operation} \label{sec:sys_design}

\begin{figure}
    \centering
    \begin{tikzpicture}[xscale=1.3]
        \coordinate (in1) at (0,1);
        \node at (.3,.2) {\Shortstack{. . .}};
        \coordinate (inK) at (0,-1);
        \node[rectangle,draw,minimum height=2.5cm,align=center] (encoder) at (1,0) {Enc.\\$E$};
        \coordinate (inBeta) at (1,{-2}-|encoder.south);
        
        \draw[->] (in1) -- (in1-|encoder.west) node[midway,above] {$\mX_1$};
        \draw[->] (inK) -- (inK-|encoder.west) node[midway,above] {$\mX_K$};
        \draw[->] (inBeta) -- (encoder.south) node[midway,right] {$\alpha_n$};
        
        \node[rectangle,draw,minimum width=2cm] (comp) at (3,0) {Comp. $H$};
        \node[rectangle,draw,minimum width=2cm] (preproc) at (3,-1) {Pre-proc.};
        \coordinate (in1') at (2.4,-2);
        \node at (3.2,-1.6) {. . .};
        \coordinate (inK') at (3.6,-2);
        
        \draw[->] (in1') -- (in1'|-preproc.south) node[midway,right] {$\mX_1$};
        \draw[->] (inK') -- (inK'|-preproc.south) node[midway,right] {$\mX_K$};
        \draw[->] (preproc) -- (comp);
        \draw[->] (encoder) -- (comp) node[midway,above] {$\tX_n$};
        
        \coordinate (out) at (5,0);
        
        \draw[->] (comp) -- (out) node[midway,above] {$\ty_n$};
    \end{tikzpicture}
    \caption{Encoding and computation operation for worker $n$.}
    \label{fig:enc_comp}
\end{figure}
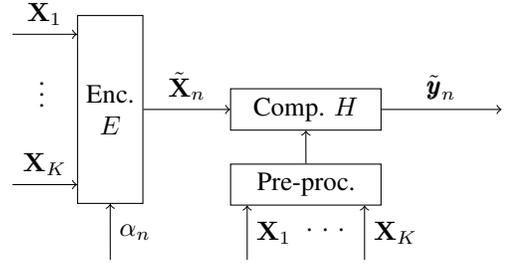

The system includes three operations: \emph{encoding}, \emph{computation}, and \emph{decoding}.
The master performs the encoding and the decoding operations while the workers do the computations.
In the following, we describe each of these operations in detail.
Figure \ref{fig:enc_comp} depicts the general structure of the encoding and computation operations for one worker.

\subsubsection*{Encoding operation}
Figure \ref{fig:encoder arch} shows the structure of the encoding function. Given an integer $G > 0$, the encoder $\opE : (\sD)^K \times \Real \to \sW$ is given by
\begin{align}
    \label{eq:encoding_poly}
    \opE(\sX, \alpha) \define \sum_{g=0}^G \mU_g\ \alpha^g.
\end{align}
The coefficients $\mU_g \in \sW$, $g=0,\dots,G$, are
obtained via the functions $\opD_g : (\sD)^K \to \sW$
such that $\mU_g \define \opD_g(\sX)$.
The functions $\opD_g$, $g=0,\dots,G$ are DNNs whose training procedure will be discussed in Section \ref{sec:training}.
Given the input set $\sX$, the function $\ope(\cdot) \define\opEX$ is a degree $G$ polynomial.
Integer $G$ is a design parameter that, in contrast to the LCC, 
does not directly depend on the number of input matrices $K$.
Encoded matrices are obtained for $N$ distinct scalars $\{\alpha_1,\dots,\alpha_N\}$ as:
\begin{align} \label{eq:encoded_matrices}
    (\forall n \in \Sseq{1}{N}) \qquad \tX_n \define \ope(\alpha_n) \in \sW,
\end{align}
where $\alpha_n \in \Real$, for all $n=1,\dots,N$.
Each encoded matrix is then sent to one of the $N$ workers for computation.

\subsubsection*{Computation operation $\opH$}
Figure \ref{fig:computer arch} shows the structure of the computation function.
Worker $n$ performs the computation operation on the encoded matrix $\tilde \mX_n$. Given an integer $P>0$, the computation operation $\opH : (\sD)^K \times \sW \to \sQ$ yields
\begin{align}
    \label{eq:computation_poly}
    \opH(\sX, \tX) \define \sum_{p=0}^P \mV_p\ \Vect(\tX^p),
\end{align}
where $\Vect(\mX)$ vectorizes matrix $\mX$ by stacking its columns.
The integer $V$ represents the number of coordinates in the range space $\sS$ of the desired function $\opf$.
The coefficients $\mV_p \in \sT$, $p=0,\dots,P$, are
obtained via DNNs $\opL_p:(\sD)^K \to \sT$ as $\mV_p \define \opL_p(\sX)$. Given coefficients $\mV_p$, we use $h(\tX_n) \define \opH(\sX, \tX_n)$ to denote the function that represents the computation operation. It takes encoded data $\tX_n$ as input and outputs the computation result $\ty_n \define \oph(\tX_n)$.  
Notice that for a given dataset $\sX$, the operation $\opH$ is common to all the workers.
Hence, the coefficients $\{\mV_p\}$ can be computed by the master, or by any one of the workers, and broadcasted to all workers.
The integer $P>0$ is a design parameter that does not directly depend on the function $\opf$.

\subsubsection*{Decoding process}
Given a scalar $\alpha \in \Real$ and a corresponding encoded data vector $\tX$, worker output $\ty=h(\tX)=h(e(\alpha))$ can be viewed as a polynomial evaluated at the point $\alpha$, since
\begin{align} \label{eq:composite}
    \oph(\ope(\alpha)) = \sum_{p=0}^P \mV_p\ \Vect\left[\left(\sum_{g=0}^G \mU_g\ \alpha^g \right)^p\right]
        = \sum_{l=0}^{GP} \vz_l\ \alpha^l,
\end{align}
where the last equality follows by taking scalars of the form $\alpha^{gp}$ out of the $\Vect{\cdot}$ operation.
Hence, the function $\oph \circ \ope$ is a polynomial of degree $G P \equiv \text{deg}(\ope) \text{deg}(\oph)$.
In \eqref{eq:composite}, the coefficients $\vz_l$, $l = 0,\dots,GP$, belong to $\sQ$.
Therefore, the number of data points required to recover $\opv \define \oph \circ \ope$ is $R = GP + 1$.
The recovery can be performed, for example, using Lagrange interpolation.
This integer $R$ is the recovery threshold of the proposed \psc~scheme.

To retrieve the final results, the master evaluates $\opv$ at scalars $\beta_1, \dots, \beta_K$, which are distinct real numbers that are chosen in the training phase,
and fixed for the given problem.
This aspect is different from LCC where these scalars can be chosen freely by the master at the time of encoding.
The final result of the decoding process is given by
\begin{align} \label{eq:desired}
    (\forall k\in\Sseq{1}{K}) \qquad \hf_k \define \oph(\ope(\beta_k)) \in \sQ.
\end{align}

The master obtains the results $\{\hf_1, \dots, \hf_K\}$ as approximations for $\{\opf(\mX_1), \dots, \opf(\mX_K)\}$.
The accuracy of this approximation, measured based on a cost function, 
is assured by the training procedure described in the next section.

\subsection{Training}
\label{sec:training}

The encoding and decoding operations in the \psc~scheme described in Section \ref{sec:sys_design}
consist of DNNs $\{\opD_0, \dots, \opD_G, \opL_0, \dots, \opL_P\}$, each with learnable parameters.
These parameters are trained using a training dataset such that the cost
of approximating $\opf(\mX_k)$ with $\hf_k$ is minimized for all $k\in\Sseq{1}{K}$ and all input datasets.
Distinct real-valued scalars $\beta_k$, $k=1,\dots,K$, are chosen at the beginning.
They remain fixed during the entire training and operation phase.
The choice of cost function depends on the application,
but it should be continuous and differentiable with respect to the learnable parameters appearing in the DNN functions.
However, one can select the specific architecture of the DNNs and the training procedure freely. We discuss these details  in Section \ref{sec:simulations} for various functions.

\begin{figure}
    \centering
    \begin{tikzpicture}
        \node (in1) at (0,3) {$\mX_1$};
        \node (in2) at (0,2) {$\mX_2$};
        \node at (0,1) {\Shortstack{. . .}};
        \node (inK) at (0,0) {$\mX_K$};
        
        \node[rectangle,draw] (nn1) at (2,3) {$\opD_0$};
        \node[rectangle,draw] (nn2) at (2,2) {$\opD_1$};
        \node at (2,1) {\Shortstack{. . .}};
        \node[rectangle,draw] (nnG) at (2,0) {$\opD_G$};
        
        \foreach \x in {in1,in2,inK} {
            \foreach \y in {nn1,nn2,nnG} {
              \draw[->] (\x) -- (\y);
            }
        }
        
        \node[circle,draw,inner sep=.1pt] (mult1) at (4,2) {$\times$};
        \node[circle,draw,inner sep=.1pt] (multG) at (4,0) {$\times$};
        \node (scalar) at (4,1.2) {$\alpha_n$};
        \node (scalarG) at (4,-.8) {$\alpha_n^G$};
        
        \draw[->] (nn2) -- (mult1) node[midway,above] {$\mU_1$};
        \draw[->] (nnG) -- (multG) node[midway,above] {$\mU_G$};
        \draw[->] (scalar) -- (mult1);
        \draw[->] (scalarG) -- (multG);
        
        \node[circle,draw,inner sep=.1pt] (add) at (5,2) {$+$};
        
        \draw[->] (nn1) -- (mult1.west|-nn1) node[midway,above] {$\mU_0$} -- (add|-nn1) -- (add);
        \draw[->] (mult1) -- (add);
        \draw[->] (multG) -- (multG-|add) -- (add);
        
        \node (out) at (6.5,2) {$\tX_n$};
        
        \draw[->] (add) -- (out);
    \end{tikzpicture}
    \caption{Encoding Function at the master.} 
    \label{fig:encoder arch}
\end{figure}
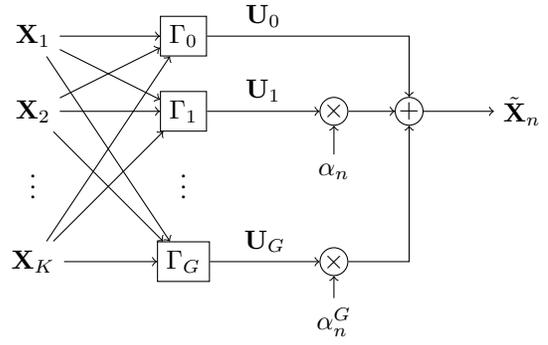

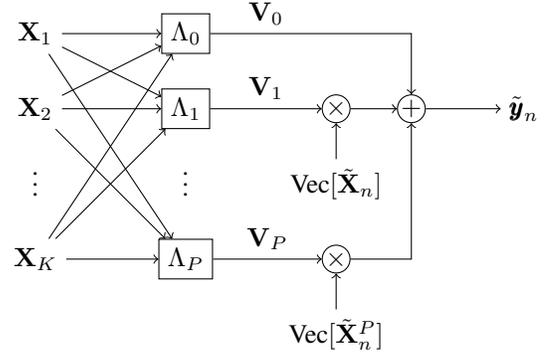
\begin{figure}
    \centering
    \begin{tikzpicture}
        \node (in1) at (0,3) {$\mX_1$};
        \node (in2) at (0,2) {$\mX_2$};
        \node at (0,1) {\Shortstack{. . .}};
        \node (inK) at (0,0) {$\mX_K$};
        
        \node[rectangle,draw] (nn1) at (2,3) {$\opL_0$};
        \node[rectangle,draw] (nn2) at (2,2) {$\opL_1$};
        \node at (2,1) {\Shortstack{. . .}};
        \node[rectangle,draw] (nnG) at (2,0) {$\opL_P$};
        
        \foreach \x in {in1,in2,inK} {
            \foreach \y in {nn1,nn2,nnG} {
              \draw[->] (\x) -- (\y);
            }
        }
        
        \node[circle,draw,inner sep=.1pt] (mult1) at (4,2) {$\times$};
        \node[circle,draw,inner sep=.1pt] (multG) at (4,0) {$\times$};
        \node (scalar) at (4,1) {$\Vect[\tX_n]$};
        \node (scalarG) at (4,-1) {$\Vect[\tX_n^P]$};
        
        \draw[->] (nn2) -- (mult1) node[midway,above] {$\mV_1$};
        \draw[->] (nnG) -- (multG) node[midway,above] {$\mV_P$};
        \draw[->] (scalar) -- (mult1);
        \draw[->] (scalarG) -- (multG);
        
        \node[circle,draw,inner sep=.1pt] (add) at (5,2) {$+$};
        
        \draw[->] (nn1) -- (mult1.west|-nn1) node[midway,above] {$\mV_0$} -- (add|-nn1) -- (add);
        \draw[->] (mult1) -- (add);
        \draw[->] (multG) -- (multG-|add) -- (add);
        
        \node (out) at (6.5,2) {$\ty_n$};
        
        \draw[->] (add) -- (out);
    \end{tikzpicture}
    \caption{Computation Function of worker $n$.} 
    \label{fig:computer arch}
\end{figure}

Figures \ref{fig:encoder arch} and \ref{fig:computer arch} respectively show the input-output structure of the
encoding and computation operations with corresponding DNNs.
The training consists of forward and backward passes over the system to update the parameters using the stochastic gradient descent (SGD) algorithm.
In each forward pass, given a dataset $\sX$ and a scalar $\beta_k$,
the sequence of encoding, computation, and decoding operations result in the output $\hf_k = \oph(\ope(\beta_k))$.
Then, in the backward pass, the parameters are updated based on the gradients of the cost w.r.t.~the parameters,
evaluated using the output $\hf_k$ and the target value $\vf_k \define \Vect(\opf(\mX_k))$.

One instance of the training data consists of inputs $\Sseq{\mX_1}{\mX_K}$, 
distinct scalars $\Sseq{\beta_1}{\beta_K}$, 
and corresponding target labels $\Sseq{\vf_1}{\vf_K}$.
The inputs $\Sseq{\mX_1}{\mX_K}$ corresponding to every training instance are sampled i.i.d.~from 
a given distribution which is representative of the true distribution of inputs to the function $\opf$.
One training step consists of forward and backward pass over a \emph{batch} of data,
and an \emph{epoch} includes multiple batches.
The SGD algorithm updates the weights after every epoch,
and the entire training runs for multiple epochs.

\section{Simulations}
\label{sec:simulations}

In this section, we apply the proposed \psc~scheme for coded computation of four frequently-used matrix functions in digital signal processing applications. In the following, we present details of those functions and elaborate on the data generation process. We also discuss relevant implementation issues highlighting unique aspects related to the design and training of DNNs. Numerical results appear at the end, accompanied by a discussion on the effect of various design parameters on the performance.

\subsection{Applied problems}
We consider the problems of computation of
($P_1$) the \emph{eigenvalues}, 
($P_2$) the \emph{eigenvector corresponding to the largest absolute eigenvalue},
($P_3$) the \emph{matrix exponential}, and
($P_4$) the \emph{determinant} of real-valued square matrices.
Each problem $P_i$, for $i=1,\dots,4$, is denoted by the function $\opfi: \sD_i \to \sS_i$, respectively.

In the training phase, the scheme learns to solve a regression problem for function approximation using a supervised learning approach.
In this framework, functions with an unbounded range are generally difficult to learn. Thus, for the present simulations, we design the input distribution for each problem to obtain a bounded range for the functions. Nevertheless, in principle, the proposed scheme applies to any function with an arbitrary domain and range
for which a system with the structure described in Section \ref{sec:AICC} is learnable to approximate the function with the desired accuracy.

For problem $P_1$, we generate an input sample as $\mX \define \frac{1}{2}(\mA + \mA^T)$, where $\mA$ is drawn uniformly from $[-1, 1]^{M \times M}$. 
For problem $P_2$, we similarly generate an input sample as $\mX \define \frac{1}{2}(\mA + \mA^T)$; however, here, $\mA$ is drawn uniformly from $[0, 1]^{M \times M}$. 
We make the value of $\opf^{(2)}$ unique by normalizing the output eigenvector to a vector of unit norm and ensuring that the first nonzero component is positive. 
For problem $P_3$, the input sample is generated as $\mX \define \mA/\normOp{\mA}$ with $\mA$ drawn uniformly from $[0, 1]^{M \times M}$. 
In addition to keeping the value of $\opf^{(3)}$ bounded, 
this ensures the numerical stability of the algorithm used to compute the ground truth in training \cite{al2010new}. Here, $\normOp{\cdot}$ denotes the operator norm of matrices.

The matrix determinant in $P_4$ is bounded by generating diagonally-dominant matrices.
More precisely, we generate the input matrices as $\mX \define \mI - \mE$, where $\mI$ is the identity matrix, and $\mE$ has zeros on the diagonal and non-diagonal elements sampled uniformly from $[-2.0/M, 2.0/M]$ \cite{ostrowski1938approximation,BRENT201521}. The number of coordinates in the range spaces $\sS_i$, for $i=1, \dots, 4$,
are, respectively, $M, M, M^2$ and $1$. For simulations, we compute the true values of the functions by using the Python's Numpy and Scipy libraries. For problems $P_1$, $P_3$, and $P_4$, we use the cost function $\opC(\hf, \vf) := \norm{\hf - \vf}$ for training, where $\vf \define \Vect(\opf(\mX))$. We use $\norm{\cdot}$ to denote the Frobenius norm for matrices and vector norm for vectors. For the eigenvector problem $P_2$, we add an additional term in the cost function to influence the network in learning a vector of unit norm, i.e.,~$\tilde{\opC}(\hf, \vf) := \opC(\hf, \vf) + 5\ (\norm{\hf} - 1.0)^2$.

\subsection{Implementation details}
Below we explicitly describe the DNN architecture and system design. For simplicity, we implement the same system design and architecture for all problems. Since the output dimension is different for each one, we choose the free parameter $V$ (cf.~\eqref{eq:computation_poly}) to match the output dimension.
The functions $\{\opD_0, \dots, \opD_G\}$ of the encoding polynomial $\ope$ are fully connected DNNs with $L=2$ hidden layers, each with $N=100$ nodes, and the same non-linear activation.
For the computation polynomial $\oph$, the DNN $\opL_0$ has the same architecture as $\opD_0$,
and $\{\opL_1, \dots, \opL_P\}$ are linear weight matrices.
Note that the master only needs to broadcast the coefficient $\mV_0 = \opL_0(\sX)$
to the workers instead of the entire input dataset $\sX$, hence saving communication resources.

As described in Section \ref{sec:AICC}, we generate the training datasets randomly for the given problem at every training iteration. This pervents overfitting of the system to a specific dataset. A single step of training consists of a batch of sets $\sX$, each with $K$ i.i.d. matrices. Each training epoch includes multiple batches.
The scalar values $\{\beta_1, \dots, \beta_K\}$ are fixed to $\beta_k = k/K$, for $k=1, \dots, K$. We implement the training using Tensorflow's (version 2.8.0) Keras module \cite{tensorflow2015}.
We employ the standard Adam optimizer from Keras, which uses an exponentially decaying learning rate.

In the testing phase, we load the model with the learned weights of DNNs corresponding to the training epoch with minimum loss. Besides, we generate the test input data in the same way as for training,
except that $R$ (recovery threshold) scalars $\{\alpha_1, \dots, \alpha_R\}$ are chosen
as distinct, but otherwise arbitrary real numbers.
In simulations, we obtain these values as $\alpha_n = n/(R+1)$.
The master first generates the coefficients of the encoder and computation polynomials (only $\mV_0$)
using the inputs $\sX$, and generates encoded matrices $\{\ope(\alpha_r)\}_{r=1}^R$.
To imitate the erasure effect of the channel, we only use the outputs of computation over these $R$ encoded matrices (corresponding to the recovery threshold).
In the final step, the master interpolates the polynomial $\opv = \oph \circ \ope$ using $R$ worker results,
and approximates the desired results as $\hf_k \define \opv(\beta_k) \approx \vf_k$ for $k=1, \dots, K$.
Error in function approximation is evaluated in terms of the normalized root mean square error (NRMSE), given by:
\begin{align} \label{eq:metric}
    \frac{\norm{\opfh_k - \opf(\mX_k)}}{\norm{\opf(\mX_k)}}.
\end{align}

Simulations are performed on a \emph{high performance computing} (HPC) cluster with nodes equipped with 
\emph{graphical processing units} (GPUs).
A GPU, compared to a \emph{central processing unit} (CPU), 
is more efficient in parallel processing of large amount data due to its highly parallelized structure.
Every HPC cluster node is equipped with NVIDIA Tesla V100 SXM2 GPU with 32Gb RAM,
and Intel Xeon(R) Gold 6150 CPU @ 2.70 GHz with 512Gb RAM.
The script used for simulations is made available for reproducibility.\footnotemark

\footnotetext{\textit{https://github.com/navya-xx/AICC.git}}

\subsection{Results}
The errors in approximate coded computation of the problems $P_1, \dots, P_4$ are listed in Table \ref{tab:errors} below.
The results presented here are evaluated on matrices of dimension $M=50$ with $K=3$ number of inputs in each dataset and recovery threshold $R=5$.
This shows the ability of the \psc~scheme to evaluate the functions $\opfi, i=1,\dots,4$, approximately in a distributed manner through coded computation.

\begin{table}[!thb]
    \centering
    \caption{Function approximation error for \psc.}
    \label{tab:errors}
    \begin{tabular}{|c|c|c|c|c|}
        \hline
            Problem $\rightarrow$ & \textbf{eigenvalue} & \textbf{eigvector} & \textbf{exponential} & \textbf{determinant}  \\
        \hline
            NMRSE (\%)    & $4.64$\% & $5.81$\% & $7.85$\% & $1.50$\% \\
        \hline
    \end{tabular}
\end{table}

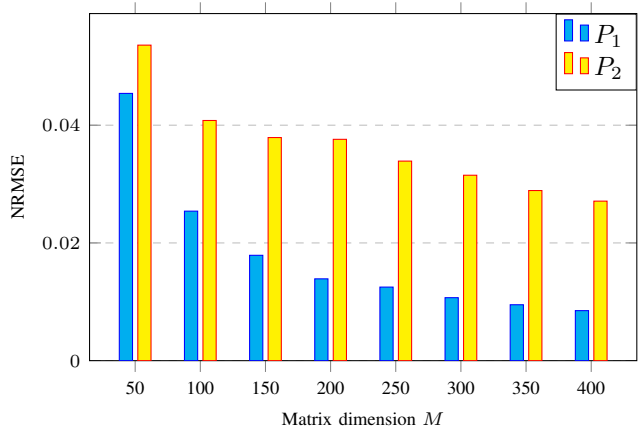
\begin{figure}[!thb]
    \centering
    \begin{tikzpicture}[scale=1.0]
        \begin{axis}[
            height=0.7\columnwidth,
            width=\columnwidth,
            ybar,
            bar width=5pt,
            ymin=0,
            xlabel={Matrix dimension $M$},
            ylabel={NRMSE},
            xtick={50, 100, 150, 200, 250, 300, 350, 400},
            xticklabels={50, 100, 150, 200, 250, 300, 350, 400},
            scaled ticks=false,
            tick label style={/pgf/number format/fixed},
            every tick label/.append style={font=\scriptsize},
            label style={font=\scriptsize},
            ymajorgrids=true,
            grid style=dashed,
            legend style={at={(1,1)}, anchor=north east},
            ]
            \addplot [draw=blue, fill=cyan] coordinates {(50, 0.0454)(100, 0.0254)(150, 0.0179)(200, 0.0139)(250, 0.0125)(300, 0.0107)(350, 0.0095)(400, 0.0085)};
            \addplot [draw=red, fill=yellow] coordinates {(50, 0.0536)(100, 0.0408)(150, 0.0379)(200, 0.0376)(250, 0.0339)(300, 0.0315)(350, 0.0289)(400, 0.0271)};
            \legend{
                $P_1$,
                $P_2$
                }
        \end{axis}
    \end{tikzpicture}
    \caption{
        Approximation error for problems $P_1$ and $P_4$ for different matrix dimensions $M$.    
            }
    \label{fig:accuracy}
\end{figure}

Figure \ref{fig:accuracy} shows the results for problems $P_1$ and $P_2$ in terms of normalized root mean square error (NRMSE)
for input matrices of dimensions $M$ varying from $50$ to $400$.
The approximation error is low ($<5$\%), and it is interesting to observe that the error decreases with increasing dimensions $M$, offering a reasonable accuracy for high dimensional problems.
Note that, with the present structure, the number of learnable parameters in the system grows at least at $\mathcal{O}(KM^2)$.
While this increases the system's capability to learn high-dimensional problems with even higher accuracy,
it comes at the price of a longer and more complex training process.

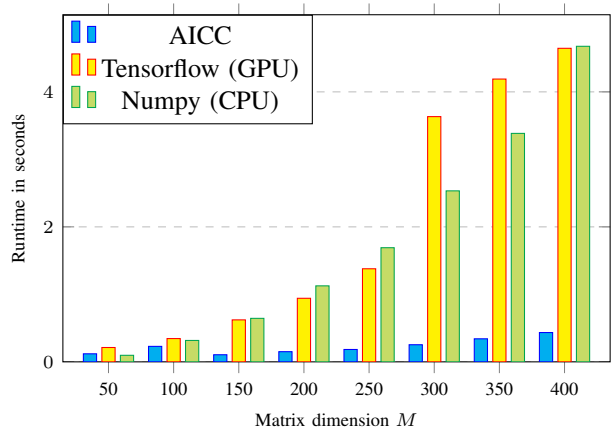
\begin{figure}[!thb]
    \centering
    \begin{tikzpicture}[scale=1.0]
        \begin{axis}[
            height=0.7\columnwidth,
            width=\columnwidth,
            ybar,
            bar width=5pt,
            ymin=0,
            xlabel={Matrix dimension $M$},
            ylabel={Runtime in seconds},
            xtick={50, 100, 150, 200, 250, 300, 350, 400},
            xticklabels={50, 100, 150, 200, 250, 300, 350, 400},
            scaled ticks=false,
            tick label style={/pgf/number format/fixed},
            every tick label/.append style={font=\scriptsize},
            label style={font=\scriptsize},
            ymajorgrids=true,
            grid style=dashed,
            legend style={at={(0,1)}, anchor=north west},
            ]
            \addplot [draw=blue, fill=cyan, discard if not={problem}{eigenvalue}] table[x=matrix_dims, y=mean_runtime, col sep=comma, row sep=newline]{runtime_results_eig.csv};
            \addplot [draw=red, fill=yellow, discard if not={run_ids}{EIG0}] table[x=matrix_dims, y=mean_runtime, col sep=comma, row sep=newline]{runtime_results_eig.csv};
            \addplot [draw=Green, fill=SpringGreen, discard if not={run_ids}{EIG1}] table[x=matrix_dims, y=mean_runtime, col sep=comma, row sep=newline]{runtime_results_eig.csv};
            \legend{
                \psc,
                Tensorflow (GPU),
                Numpy (CPU)
                }
        \end{axis}
    \end{tikzpicture}
    \caption{
        Comparison of runtime of \psc~and standard implementations for computing eigenvalues.    
            }
    \label{fig:runtime}
\end{figure}

In Figure \ref{fig:runtime}, we compare the runtime of the proposed \psc~scheme with
standard implementations in Python (called STD for brevity) of the problem $P_1$
for exact computation of eigenvalues of a matrix.
The input dataset consists of 128 batches of $K=3$ matrices, and runtime is averaged over 1000 runs.
The computation task corresponding to each scheme is performed on a single node in the cluster.
The runtime of STD is reported for both CPU and GPU based implementations,
using the Numpy function \texttt{np.linalg.eigvalsh}
and the Tensorflow function \texttt{tf.linalg.eigvalsh}, respectively.
The \psc~scheme runs all master and worker computations on a single GPU, 
so that it is comparable to the STD GPU computations 
in terms of computational resources used per unit of time.
As $M$ grows, the runtime of the STD scheme increase much faster than that of \psc.

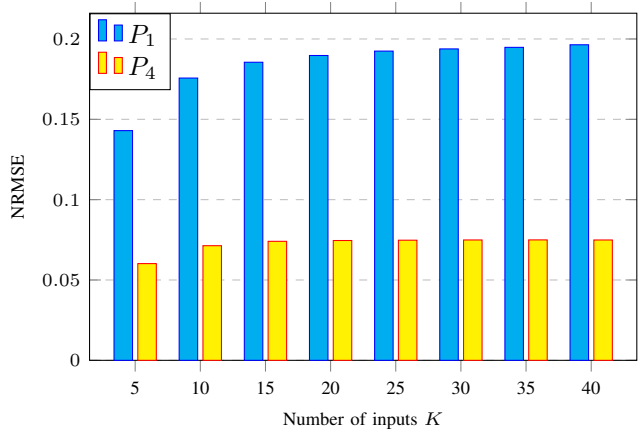
\begin{figure}[!thb]
    \centering
    \begin{tikzpicture}[scale=1.0]
        \begin{axis}[
            height=0.7\columnwidth,
            width=\columnwidth,
            ybar,
            bar width=7pt,
            ymin=0,
            xlabel={Number of inputs $K$},
            ylabel={NRMSE},
            xtick={5, 10, 15, 20, 25, 30, 35, 40},
            xticklabels={5, 10, 15, 20, 25, 30, 35, 40},
            scaled ticks=false,
            tick label style={/pgf/number format/fixed},
            every tick label/.append style={font=\scriptsize},
            label style={font=\scriptsize},
            ymajorgrids=true,
            grid style=dashed,
            legend style={at={(0,1)}, anchor=north west},
            ]
            \addplot [draw=blue, fill=cyan, discard if not={problem}{eigenvalue}] table[x=num_inputs, y=mean_err_interp, col sep=comma]{numinputs.csv};
            \addplot [draw=red, fill=yellow, discard if not={problem}{matrix_det}] table[x=num_inputs, y=mean_err_interp, col sep=comma]{numinputs.csv};
            \legend{
                $P_1$,
                $P_4$
                }
        \end{axis}
    \end{tikzpicture}
    \caption{Varying number of inputs $K$ at fixed recovery threshold $R=5$.}
    \label{fig:num_inputs}
\end{figure}

In Figure \ref{fig:num_inputs}, we compare the performance of $P_1$ and $P_4$ with an increasing number of inputs $K$,
while keeping the recovery threshold of the scheme constant at $R=5$ and the matrix dimension constant at $M=10$.
Notice that the approximation error does not increase too quickly with increasing $K$.
Therefore, for applications with a large number of input matrices needing to be processed at a time,
\psc~enables approximate recovery with only a small number of worker results.

In further experiments performed over matrix dimensions $M=10$ and $50$, 
increasing the recovery threshold $R$ showed a slight downward trend in approximation error
for small $R$ as expected, but reaching an error floor quickly.
Further investigation is required in this regard, but one possible explanation is that
the learning saturates for the analyzed problems too soon to show the expected trend.

\clearpage
\bibliographystyle{IEEEtran}
\bibliography{references}

\end{document}